\documentclass[twocolumn, preprintnumbers, amsmath, amssymb, showpacs]{revtex4}
\usepackage{graphicx}
\usepackage{dcolumn}
\usepackage{bm}

\def\be{\begin{equation}}
\def\ee{\end{equation}}
\def\bd{\begin{displaymath}}
\def\ed{\end{displaymath}}
\def\-{\phantom{-}}

\begin{document}
\title{Dynamics of Bulk vs. Nanoscale WS$_2$: Local Strain and Charging Effects}

\author{R.D. Luttrell$^1$, S. Brown$^1$, J. Cao$^1$, J.L. Musfeldt$^1$, R.
Rosentsveig$^2$, and R. Tenne$^2$}

\affiliation{$^1$Department of Chemistry, University of Tennessee,
Knoxville, Tennessee, USA 37996}

\affiliation{$^2$Department of Materials and Interfaces, Weizmann
Institute of Science, Rehovot, Israel 76100}

\begin{abstract}

We measured the infrared vibrational properties of bulk and
nanoparticle WS$_2$ in order to investigate the structure-property
relations in these novel materials. In addition to the
symmetry-breaking effects of local strain, nanoparticle curvature
modifies the local charging environment of the bulk material.
Performing a charge analysis on the \emph{xy}-polarized E$_{1u}$
vibrational mode, we find an approximate 1.5:1 intralayer charge
difference between the layered 2H material and inorganic
fullerene-like (IF) nanoparticles. This effective charge difference
may impact the solid-state lubrication properties of nanoscale metal
dichalcogenides.

\end{abstract}

\pacs{63.22.+m, 78.20.Ci, 61.46.+w, 78.30.-j}

\narrowtext

\maketitle

\clearpage

\section{Introduction}

\begin{figure}
\vspace{0.1cm}
\includegraphics[width = 3.5 in] {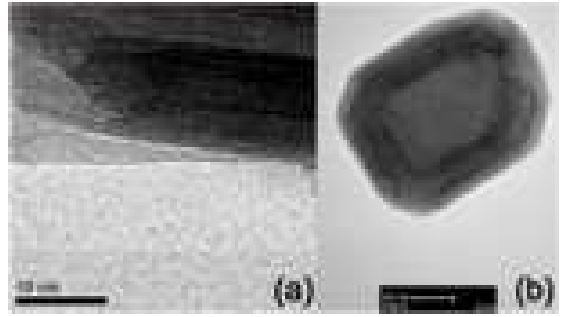}
\caption{\label{fig_ws2nanoparticleb} Transmission electron
microscope images of (a) layered 2H-WS$_2$  and (b) a representative
IF-WS$_2$ nanoparticle (b). Each nanoparticle consists of a hollow
core and several W-S-W layers. The average particle diameter is
$\sim$120 - 200 nm for the materials of interest here. The IF
nanoparticle shown here is slightly smaller than average.}
\end{figure}

Inorganic fullerene-like (IF) nanostructures 
have recently attracted attention due to their unique closed cage
structures  and outstanding solid-state lubricating behavior
\cite{Tenne1992,Margulis1993,Parilla1999}. Just as carbon fullerenes
are nanoscale analogs of layered graphite, IF nanoparticles and
nanotubes are curved analogs of the corresponding
quasi-two-dimensional material.
Layered and nanoscale metal dichalcogenides are prototypes in this
regard, and
the discovery of the WS$_2$-based family of IF nanoparticles
(Fig.~\ref{fig_ws2nanoparticleb}) provides the opportunity to
investigate structure-property relations in bulk vs. nanoscale
materials.  At the same time, the IF materials hold out the
potential for important applications. In addition to use in
rechargeable batteries, optical devices, and in impact-resistant
nanocomposities, extensive mechanical properties testing
demonstrates that the friction coefficient of IF-WS$_2$
nanoparticles is reduced up to 50\% compared with the 2H-WS$_2$
parent compound, maintaining excellent lubricating behavior even
under very high loads, ultra-high vacuum, and in humid conditions
\cite{Rapoport1997,Rapoport1999,Rapoport2003,Chen2002,Schuffenhauer2004,Rapoport2005,Zhu05}.
Because of these observations, major efforts have been directed at
understanding the connection between bulk and microscopic properties
and exploiting the commercial promise of these novel nanomaterials.

\begin{figure}
\includegraphics[width = 3.4 in]{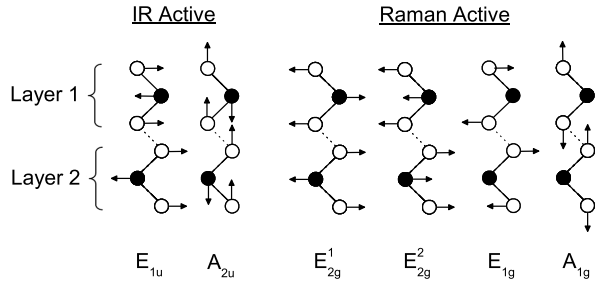}
\caption{\label{fig_ws2mode} Displacement vectors of the infrared
and Raman active modes of 2H-WS$_2$. The dashed line between the two
layers represents the weak van der Waals force.}
\end{figure}

2H-WS$_2$ belongs to the space group P6$_3$/mmc ($D_{6h}^4$) and
contains two formula units per unit cell \cite{2HNote}. The bonding
is well-known to consist of strong covalent intralayer forces and
weak van der Waals interactions between adjacent layers
\cite{Wilson1969,Schutte1987}. Each MX$_2$ layer (M = group VIB
metal, X = group VIA element) contains a layer of metal atoms,
sandwiched between two chalcogen layers, with each metal atom bonded
to six chalcogen atoms in a trigonal prismatic arrangement. A group
theoretical analysis (Table~\ref{tab_ws2group}, Appendix) gives a
total of 18 normal modes \cite{Verble1970}. The doubly degenerate
E$_{1u}$ and singly degenerate A$_{2u}$ vibrational modes are
infrared active; the conjugate gerade 
modes are Raman active. Figure \ref{fig_ws2mode} displays the
relevant
displacement vectors. 
Note that the infrared active \emph{xy}-polarized E$_{1u}$ and
\emph{z}-polarized A$_{2u}$ modes are associated with intralayer and
interlayer motion, respectively. This directional selectivity
provides a sensitive and microscopic probe of charge and bonding
interactions that we will employ in this work to assess effective
total and local charge differences between the bulk and nanoscale
materials
\cite{Uchida1978,Lucovsky1973,Lucovsky1971,Lucovsky1976,Burstein1970}.
X-ray diffraction reveals that the local structure of 2H-WS$_2$ is
preserved within the unit cell of an IF nanoparticle with the
exception of a 2\% lattice expansion along the \emph{z}-axis
\cite{Feldman1995,Feldman1996}. This lattice expansion is attributed
to strain in the curved WS$_2$ layers
\cite{Feldman1996,Srolovitz1995}, each of which has a slightly
different radius due to layer inhomogeneities
(Fig.~\ref{fig_ws2nanoparticleb}(b)). The lattice modes of 2H-WS$_2$
were previously investigated using combined Raman and inelastic
neutron scattering, demonstrating that the two-phonon resonance
Raman effects are second-order processes involving the longitudinal
acoustic mode at the K point of the Brillouin zone
\cite{Sourisseau1991}. Previous studies also indicate that the new
Raman peaks in spectra of IF-WS$_2$ nanotubes and nanoparticles
should be assigned as disorder-induced zone edge
phonons \cite{Frey1998b}. 
More recent Raman work demonstrates that the 33 cm$^{-1}$ E$_{2g}$
shearing or ``rigid layer" mode is almost completely blocked in
IF-WS$_2$ due to surface strain hindering intralayer motion in the
nanoparticles \cite{unpublished}. Extensive optical properties and
STM work indicates that the indirect gap is smaller in the IF
nanoparticles compared to that in the bulk
\cite{Frey1998,Hershfinkel1994}. A and B exciton positions are also
sensitive to confinement and the number of layers in the
nanoparticle \cite{Frey1998b,Frey1998}.

In order to investigate structure-property relationships in these
chemically identical but morphologically different metal
dichalcogenides, we measured the infrared reflectance spectra of
both 2H- and IF-WS$_2$ powders and performed a charge analysis to
extract total and local effective charge from the oscillator
parameters of the major infrared active phonon modes. We found an
approximate 1.5:1 intralayer charge difference between the 2H- bulk
and IF-nanoparticle materials,
respectively. 
The trend is different in the interlayer direction, reflecting a
slightly enhanced $z$ interaction
in the nanoparticles. We discuss how differences in both charge and
strain may be connected to the macroscopic properties of these
materials.

%

\section{Experimental Methods}

IF-WS$_2$ was prepared from its oxide precursor, WO$_3$, following
previously published procedures \cite{Tenne1992,Feldman1996,
Homyonfer1997,Feldman2000, Rosentsveig2002, Margolin2004,
Tenne2003}. The IF nanoparticles of interest in this work range in
size from $\sim$120 - 200 nm in diameter. The particle size, shape
and distribution have been studied by x-ray powder diffraction
\cite{Feldman1996}, scanning tunneling microscopy
\cite{Hershfinkel1994}, and high-resolution transmission electron
microscopy \cite{Hershfinkel1994,Homyonfer1997,Feldman1996}.
Pressed isotropic pellets were prepared to investigate the dynamical
properties. 
Bulk 2H-WS$_2$ was also measured for comparison (Alfa Aesar,
99.8\%).

Near-normal infrared reflectance was measured over a wide frequency
range using a series of spectrometers including a Bruker 113V
Fourier transform infrared spectrometer, an Equinox 55 Fourier
transform instrument (equipped with a microscope attachment), and a
Perkin Elmer Lambda 900 grating spectrometer, covering the frequency
range from 25 - 52000 cm$^{-1}$.  A helium-cooled bolometer detector
was employed in the far-infrared for added sensitivity. Both 0.5 and
2 cm$^{-1}$ resolution were used in the infrared, whereas 3 nm
resolution was used in the optical regime. Variable temperature
measurements were carried out with a continuous-flow helium cryostat
and temperature controller. A Kramers-Kronig analysis was used to
calculate the optical constants from the measured reflectance,
yielding information on the dispersive and lossy response of each
material \cite{Wooten1972,optconst}. Standard peak-fitting
techniques were employed, where appropriate.

\section{Results and Discussion}

\subsection{Understanding Charge Localization Effects in 2H- and IF-WS$_2$}

\begin{figure}
\includegraphics[width = 3.2 in]{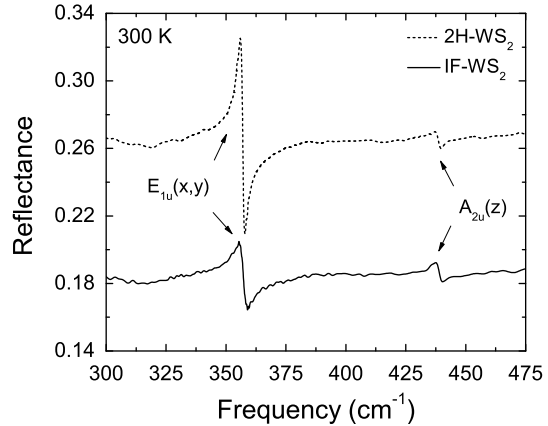}
\caption{\label{fig_ws2300Kref} 300 K reflectance spectra of 2H- and
IF-WS$_2$. The E$_{1u}$ and A$_{2u}$ modes are infrared active in
these materials, with $xy$- and $z$-directed polarizations,
respectively. }
\end{figure}

Figure~\ref{fig_ws2300Kref} displays a close-up view of the far
infrared reflectance spectra of 2H- and IF-WS$_2$ at 300 K. We
assign the major features in the reflectance (at 356 and 437
cm$^{-1}$) as the E$_{1u}$ and A$_{2u}$ modes, respectively
\cite{Verble1970,conductivitynote}. These modes are strikingly
different in the two materials.  The E$_{1u}$ mode appears damped
and suppressed in IF-WS$_2$ compared to the 2H- analog, whereas the
A$_{2u}$ mode is slightly more pronounced in the IF compound
compared to that in the bulk. These differences can be quantified
using traditional dielectric oscillator models and fitting
techniques which, in combination with appropriate models, allow us
to assess the charge characteristics of the nanomaterial as compared
to the bulk.

\begin{table}
\caption{\label{tab_ws2oscillator}} Classical Oscillator Parameters
and Optical Phonon Frequencies for 2H- and IF-WS$_2$ \\
\begin{ruledtabular}
\begin{tabular}{ccccc}
& \multicolumn{2}{c}{E$_{1u}$ Mode} & \multicolumn{2}{c}{ A$_{2u}$ Mode}\\

300 K & 2H-WS$_2$  & IF-WS$_2$ & 2H-WS$_2$ & IF-WS$_2$\\
 \hline
Oscillator strength, S & 0.031 & 0.014 & 0.002 & 0.002\\
Damping constant, $\gamma$ & 0.005 & 0.010 & 0.004 & 0.006\\

$\omega_{TO}$ (cm$^{-1}$) & 356.47 & 356.97 & 438.32 & 439.01\\

$\omega_{LO}$ (cm$^{-1}$) & 357.10 & 357.28 &  & \\

$\omega_{LO}$-$\omega_{TO}$ (cm$^{-1}$) & 0.63 & 0.31 &  & \\

$\epsilon_\infty$ & 9.58 & 6.26 &&\\

\hline \hline
&&&&\\
10 K  \\
 \hline
Oscillator strength, S & 0.040 & 0.017 & 0.005 & 0.004\\

Damping constant, $\gamma$ & 0.004 & 0.010 & 0.006 & 0.004\\

$\omega_{TO}$ (cm$^{-1}$) & 358.30 & 358.61 & 440.80 & 440.87\\

$\omega_{LO}$ (cm$^{-1}$) & 359.14 & 359.20 &  & \\

$\omega_{LO}$-$\omega_{TO}$ (cm$^{-1}$) & 0.84 & 0.59 &  & \\

$\epsilon_\infty$ & 8.92 & 6.46 &&\\

\end{tabular}
\end{ruledtabular}
\end{table}

One well-established approach for quantifying charge in a material
involves assessment of both total and local effective charge
\cite{Uchida1978,Lucovsky1973,Lucovsky1971,Lucovsky1976,Burstein1970}.
As discussed by Burstein $\emph{et al.}$ \cite{Burstein1970},
macroscopic effective charge, $\emph{e}_T^*$, is a measure of the
electric moment per unit cell and contains contributions from both
the localized charge on the ion sites and the charge generated
throughout the unit cell. Therefore, $\emph{e}_T^*$, can be
separated into a localized part, $\emph{e}_l^*$, and a nonlocalized
part, $\emph{e}_{nl}^*$, \be
\emph{e}_T^*=\emph{e}_l^*+\emph{e}_{nl}^*.\ \label{nonlocal} \ee
Here, $\emph{e}_l^*$ is defined as the localized moment generated
per unit displacement of an ion. It induces a local field through
dipole-dipole interactions which contributes to the reduction of the
transverse optical (TO) phonon frequency. Although $\emph{e}_T^*$
can be a good measure of bond ionicity or covalency in layered
MX$_2$ transition-metal dichalcogenides, $\emph{e}_l^*$ gives a more
appropriate representation of bonding interactions because it
quantifies charge on the ionic sites \cite{Uchida1978}.

Total macroscopic effective charge, $\emph{e}_T^*$, is given as \be
\frac{\emph{e}_T^*}{\emph{e}}=
\frac{\omega_{TO}c}{\emph{e}}\sqrt{\frac{4\pi^2\epsilon_{\textrm{0}}\bar{m}S}{N}}.\
\label{macro} \ee Here, $\omega_{TO}$ is the TO phonon frequency (in
cm$^{-1}$), $S$ is the oscillator strength, $N$ is the number of
WS$_2$ formula units per unit volume, $\bar{m}$ is the mode mass
\cite{Herzberg1945}, $\emph{e}$ is the electronic charge, $c$ is the
speed of light (in cm/s), and $\epsilon_{\textrm{0}}$ is the
permittivity of free space. Note that for comparison of two similar
materials, such as 2H- and IF-WS$_2$, oscillator strength and  TO
phonon frequency will be the distinguishing parameters. Localized
effective charge \cite{Uchida1978}, $\emph{e}_l^*$, is determined as
\be
\frac{\emph{e}_l^*}{\emph{e}}=c\sqrt{(\omega_{LO}^2-\omega_{TO}^2)\frac{\bar{m}\epsilon_{\textrm{0}}}{\emph{e}^2LN}}
.\ \label{local} \ee Here, $\omega_{LO}$ is the longitudinal optical
(LO) phonon frequency (in cm$^{-1}$) and $L$ is the Lorentz factor
for a hexagonal lattice \cite{Mueller1936}. Precise measurement of
the LO-TO splitting is key to distinguishing  local charge
differences in 2H- and IF-WS$_2$.

\begin{figure}[t]
\includegraphics[width = 3.5 in]{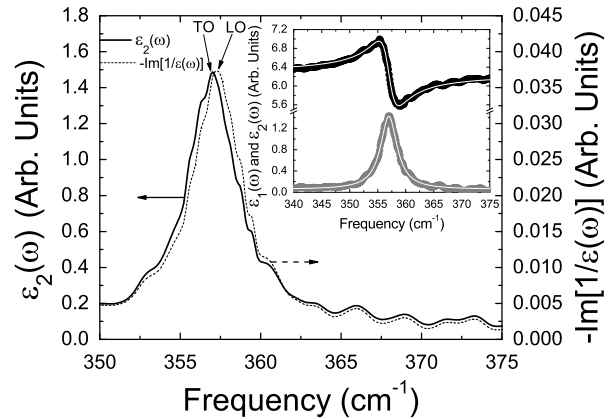}
\caption{\label{fig_iftolo} Frequency dependence of the imaginary
part of the dielectric function, $\epsilon$$_2$($\omega$), and the
energy loss function, $-$Im(1/$\epsilon$($\omega$)), of IF-WS$_2$ at
300 K. Longitudinal optical (LO) and transverse optical (TO)
frequencies are indicated. The inset displays an oscillator fit
(white line) to the real (black) and imaginary (gray) parts of the
dielectric function, $\epsilon_1({\omega})$, and
$\epsilon_2({\omega})$, for IF-WS$_2$ at
300 K. } 
\end{figure}

In order to obtain the parameters needed to extract total effective
and local charge for both 2H- and IF-WS$_2$, we carried out a
Kramers-Kronig analysis of the measured reflectance. LO and TO
phonon frequencies are obtained directly from the optical constants,
for instance, as peaks in the energy loss function,
$-$Im(1/$\epsilon$($\omega$)), and in the imaginary part of the
frequency dependent dielectric function, $\epsilon_2({\omega})$,
respectively. Figure \ref{fig_iftolo} shows an example of these
optical constants and the straightforward extraction of LO and TO
phonon frequencies for IF-WS$_2$ at 300 K.
Oscillator strength, $S$, is obtained by simultaneously fitting the
real and imaginary parts of the complex dielectric function,
$\tilde{\epsilon}$($\omega$) = $\epsilon_1$($\omega$) +
i$\epsilon_2$($\omega$), using the three parameter model \be
\tilde{\epsilon}(\omega)=\epsilon_\infty + \sum_j \frac{S_j
\omega_j^2}{\omega_j^2-\omega^2-i\gamma_j\omega_j\omega}.\
\label{fit} \ee Here, the subscript $j$ refers to the mode of
interest, $\gamma$ is the damping constant, and $\epsilon_\infty$ is
the high frequency dielectric constant
\cite{Wooten1972,Lucovsky1973,Gervais1983,Lucovsky1976}. As an
example, the inset of Fig.~\ref{fig_iftolo} shows an oscillator fit
of the real and imaginary parts of the complex dielectric function
for the E$_{1u}$ mode of IF-WS$_2$ at 300 K. The classical
oscillator parameters of both 2H- and IF-WS$_2$ are summarized in
Table~\ref{tab_ws2oscillator}. These parameters, along with  Eqns.
\ref{macro} and \ref{local}, allow us to evaluate the charge
characteristics.

\begin{table}[t]
\caption{\label{tab_ws2charge}} Macroscopic and Localized Effective Charge Values for 2H- and IF-WS$_2$ \\
\begin{ruledtabular}
\begin{tabular}{lccccc}
&& \multicolumn{2}{c}{E$_{1u}$ Mode} & \multicolumn{2}{c}{ A$_{2u}$ Mode}\\

&& 2H-WS$_2$  & IF-WS$_2$ & 2H-WS$_2$ & IF-WS$_2$\\
 \hline \hline
&&&&&\\
300 K &$\emph{e}_T^*$/$\emph{e}$ & 0.45 & 0.30 & 0.15 & 0.16\\
& $\emph{e}_l^*$/$\emph{e}$ & 0.20 & 0.14 &&\\
&&&&&\\
\hline
&&&&&\\

10 K &$\emph{e}_T^*$/$\emph{e}$ & 0.51 & 0.34 & 0.22 & 0.20\\
& $\emph{e}_l^*$/$\emph{e}$ & 0.23 & 0.20 &&\\
&&&&&\\
\end{tabular}
\end{ruledtabular}

\footnotemark {$\emph{e}_l^*$ cannot be obtained for the
\emph{z}-polarized interlayer A$_{2u}$ mode due to the negative
Lorentz factor \cite{negativeL}.}

\footnotemark {Error bars on the total and local charge values are
$\pm$ 0.02.}

\end{table}

The small LO-TO splitting in both 2H- and IF-WS$_2$ indicates that
the metal-chalcogen bond is highly covalent within the layer, in
agreement with previous results for WS$_2$ as well as other
well-known covalent compounds including MoS$_2$, MoSe$_2$, and
WSe$_2$
\cite{Burstein1970,Lucovsky1971,Lucovsky1973,Uchida1978,Lucovsky1976}.
In contrast, Uchida and Tanaka report large LO-TO splittings for
several group IV transition-metal dichalcogenides, including
1T-TiSe$_2$, 1T-ZrSe$_2$, and 1T-HfSe$_2$, which are considered to
be highly ionic materials \cite{Uchida1978}. Physically, the smaller
LO-TO splitting of the E$_{1u}$ mode in IF-WS$_2$  (Table 1),
indicates that the IF- nanoparticles are slightly more covalent than
the parent 2H- compound. This result indicates that nanoparticle
curvature changes the charge-sharing environment within the layer.

Table~\ref{tab_ws2charge} displays the total macroscopic and local
charge for both 2H- and IF-WS$_2$. Using the \emph{xy}-polarized
E$_{1u}$ mode as a probe of charge within the layer, we find that
$\emph{e}_T^*$ = 0.45 for 2H-WS$_2$ and $\emph{e}_T^*$ = 0.30 for
IF-WS$_2$ at 300 K \cite{UchidaData}. Thus, IF-WS$_2$ has
approximately two thirds the intralayer charge  as 2H-WS$_2$. This
approximate 1.5:1 charge difference is replicated in the local
charge numbers, with $\emph{e}_l^*$ decreasing proportionally in the
IF compound. The intralayer charge differences summarized in
Table~\ref{tab_ws2charge} can be traced to differences in the LO-TO
splitting and oscillator strength of the E$_{2u}$ mode in 2H- and
IF-WS$_2$. As already mentioned, these changes are easily observed
in the spectral data (Fig. \ref{fig_ws2300Kref}).
The
results imply that there is in fact a significant difference in the
local environment and chemical bonding between the 2H- and IF-
materials and that
%
nanoparticle curvature changes the charging environment within the
plane. A blocked ``rigid layer" E$_{2g}$ Raman mode is also
consistent with these observations \cite{unpublished}.

In the interlayer direction, the charge trend as characterized by
the behavior of the A$_{2u}$ mode is different
(Table~\ref{tab_ws2charge}).
At 300 K, $\emph{e}_T^*$ = 0.16 for IF-WS$_2$  compared with
$\emph{e}_T^*$ = 0.15 for 2H-WS$_2$. That $\emph{e}_T^*$ is larger
in IF-WS$_2$ is indicative of slightly stronger interlayer
interaction and enhanced charge environment in the curved
nanoparticles compared with the bulk. As expected, the total
effective charge within the layer ($\emph{e}_T^*$ from the E$_{1u}$
mode) is always larger than that between
layers ($\emph{e}_T^*$ from the A$_{2u}$ mode), 
indicating that the majority of charge resides within the metal
dichalcogenide layer in both materials.

Table~\ref{tab_ws2charge} also displays the total and local
effective charge of 2H- and IF-WS$_2$ at low temperature. Although
the exact values of total and local effective charge differ slightly
from their 300 K values (for instance, the total charge extracted
from analysis of the \emph{xy}-polarized  E$_{1u}$ mode of 2H-WS$_2$
is 0.51 at 10 K and 0.45 at 300 K), the overall trends between the
layered and nanomaterial remain similar to those discussed above.
Macroscopic effective charge within the layer decreases from 0.51 in
the bulk to 0.34 in the nanomaterial (again, an approximate  1.5:1
charge difference), and local charge makes up approximately 50\% of
the total charge. The total charge extracted from the interlayer
A$_{2u}$ mode is slightly larger in the 2H material than in the IF
nanoparticles at low temperature.

\subsection{Curvature-Induced Local Symmetry Breaking in IF-WS$_2$}

Are there other manifestations of curvature in the nanoparticles
besides the aformentioned total and local effective charge
differences?
Certainly, strain and confinement 
have been of recent interest in both vanadium oxide inorganic
nanotubes and silicon nanowires \cite{Cao04,Adu05}. In both cases,
strain broadens the vibrational modes. Another effect of curvature
is that the local, short range symmetry is formally lower than (and
a subgroup of)  the unstrained bulk. The reduction of local symmetry
can change the selection rules, allowing formerly ``infrared-silent"
modes to become infrared active \cite{jannote,Loa2001}. Further, the
curvature of each metal dichalcogenide layer within the nanoparticle
is not uniform. This inhomogeneous
structure also results in mode dispersion. 
Evidence for 
these effects, while present in the 300 K spectrum of IF-WS$_2$
(Fig. \ref{fig_ws2300Kref}), is best illustrated in the low
temperature spectral response.

\begin{figure}[t]
\includegraphics[width = 3.3 in]{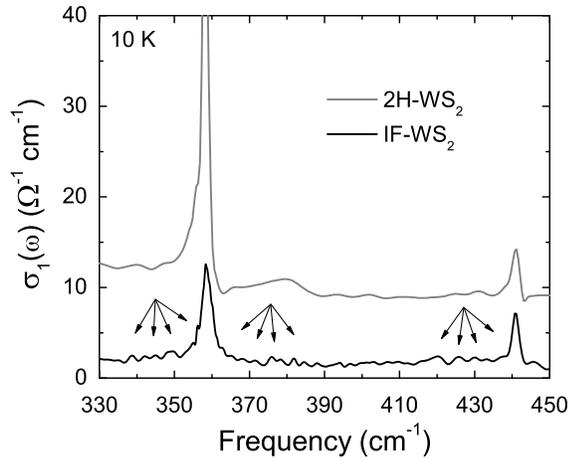}
\caption{\label{fig_WS2sigma1c} Optical conductivity of 2H- and
IF-WS$_2$ at 10 K. Arrows highlight the low-temperature symmetry
breaking in the IF material. Ref. \cite{finestructure} lists the
frequencies of these features. All of these small modes are
reproducible, although it should be noted that the exact details of
this fine structure will likely depend on particle size and
distribution within a sample, the number of nanoparticle layers, and
the nature of the defects.
}
\end{figure}

Figure \ref{fig_WS2sigma1c} displays the optical conductivity,
$\sigma_1(\omega)$, of 2H- and IF-WS$_2$ at 10 K \cite{optconst}.
The low temperature spectral response of both materials is still
dominated by the E$_{1u}$ and A$_{2u}$ modes, although because of
the additional fine structure, the nanoparticle response is clearly
much richer and more complicated than that of the bulk material. We
attribute the additional vibrational structure to the formally
``silent" and combination modes, activated (and dispersed) in the
spectrum of IF-WS$_2$ by the symmetry breaking that results from the
curved cage structure. For instance, some Raman modes are well-known
conjugates of infrared active features \cite{Sekine1980}. The
E$_{2g}^1$ and E$_{1u}^2$ conjugate pair is an example. The symmetry
analysis and vector displacement diagrams of Verble and Wieting
\cite{Verble1970} (see also Table~\ref{tab_ws2group}, Appendix) also
provide several candidates for silent mode activation, stating that
the inactive B$_{1g}$, B$_{2u}$, and E$_{2u}$ modes are nearly
degenerate with several Raman- and infrared-active modes.
Thus, the optical conductivity of IF-WS$_2$ likely contains weak
contributions from all of these first-order structures, along with a
substantial number of peaks that originate from mode combinations
and disorder effects.

Group theory can also be used to investigate possible second order
infrared active combinations. When a direct product of two
fundamental modes contains A$_{2u}$ or E$_{1u}$ symmetry in its
character, it follows that a combination of those two modes will be
infrared active. For example,
 E$_{1u}$ $\bigotimes$ E$_{2g}$ = B$_{1u}$ + B$_{2u}$ +
 E$_{1u}$, and A$_{2u}$ $\bigotimes$ E$_{1g}$ = E$_{1u}$. These
 combinations are both infrared active because their reducible representation contains
 E$_{1u}$. (Taking the direct product of this state with the $x$, $y$, and $z$ dipole moment
 operators clearly contains the totally symmetric group.)
 Similarly, the sum and difference of other fundamental vibrational
 frequencies can lead to various infrared active combinations like A$_{2u}$
$\bigotimes$ A$_{1g}$,  E$_{1u}$ $\bigotimes$ A$_{1g}$, E$_{1u}$
$\bigotimes$ E$_{1g}$, B$_{2u}$ $\bigotimes$ E$_{2g}$ and B$_{2u}$
$\bigotimes$ B$_{1g}$. We note that the longitudinal acoustic mode
is intimately and extensively involved with many combination modes
that appear in resonance Raman \cite{Sourisseau1991} and may be
involved in the infrared-active combination modes of IF-WS$_2$ as
well.

\section{Conclusion}

We report the infrared vibrational properties of bulk and
nanoparticle WS$_2$ in order to investigate the structure-property
relations in these novel materials. In addition to the
symmetry-breaking effects of local strain, nanoparticle curvature
modifies the well-known charge environment of the bulk material.
Using the E$_{1u}$ mode as a local probe of charge behavior within
the layer, we find an approximate 1.5:1 intralayer charge difference
between the  2H-  and  IF- materials. The trend is different in the
interlayer direction (probed by the A$_{2u}$ mode) at 300 K. Here,
effective charge increases slightly in the IF material compared to
that in the bulk. We attribute the  stronger interlayer interaction
to nanoparticle curvature. Elucidating the relationship between
structure and effective total and local charge in 2H- and IF-WS$_2$
is only the first step in understanding the fundamental interactions
underlying the phenomenal mechanical and solid state lubricating
properties of nanoscale metal dichalcogenides. Tuning MX$_2$ bond
covalency may, for instance, allow simultaneous exploration of
macroscopic mechanical properties, the
charge environment, 
and potential surface effects.

\section{Acknowledgements}

Work at the University of Tennessee is supported by the Materials
Science Division, Office of Basic Energy Sciences at the U.S.
Department of Energy under Grant No. (DE-FG02-01ER45885). Work at
the Weizmann Institute of Science is supported by the Helen and
Martin Kimmel Center for Nanoscale Science and by NanoMaterials,
Ltd. R.T. is the holder of the Drake Family Chair in Nanotechnology.
We thank Albert Migliori and David Tom$\acute{\textrm{a}}$nek for
interesting discussions and Ronit Popovitz-Biro for the transmission
electron microscope images.

\renewcommand{\thetable}{A.\arabic{table}}
\setcounter{table}{0}
\section*{APPENDIX: Group Theoretical Results for 2H-WS$_2$}

In 1970, Verble and Wieting carried out a complete symmetry analysis
on hexagonal layered compounds  with the goal of analyzing
vibrational mode symmetries.
Our analysis differs from that in Ref. \cite{Verble1970} in two
ways: (1) a C$_2^{\prime\prime}$ correlation table accounts for the
D$_{3h}$ 2(c) site symmetry of tungsten, and (2) a $\sigma_d$
correlation table accounts for the C$_{3v}$ 4(f) site symmetry of
sulfur \cite{Schutte1987,Fateley1972}. In Ref. \cite{Fateley1972}
refer to Table 14, p. 50, and D$_{6h}$ correlation table, p. 210.
Table \ref{tab_ws2group} summarizes our results. Local strain in the
IF-WS$_2$ breaks the selection rules of the 2H- ``parent compound",
as detailed in the text.

\begin{table}[h]
\caption{\label{tab_ws2group}}
Group Theoretical Analysis for 2H-WS$_2$ \\
\begin{ruledtabular}
\begin{tabular}{cccccl}
\multicolumn{1}{c}{Atom} && \multicolumn{2}{c}{Site Symmetry} &&
\multicolumn{1}{c}{Irreducible Representation} \\
\hline
W && ~~~D$_{3h}$ & 2(c) && A$_{2u}$ + B$_{1g}$ + E$_{2g}$ + E$_{1u}$ \\
S && ~~~C$_{3v}$ & 4(f) && A$_{1g}$ + A$_{2u}$ + B$_{1g}$ + B$_{2u}$ \\
&&&&&~~~~~+ E$_{1g}$ + E$_{1u}$ + E$_{2g}$ + E$_{2u}$ \\

\hline \hline

$\Gamma_{Total}$ &&&&&2A$_{2u}$ + 2B$_{1g}$ + A$_{1g}$ + B$_{2u}$ \\
&&&&&~~+ E$_{1g}$ + 2E$_{2g}$ + 2E$_{1u}$ + E$_{2u}$ \\

\hline

$\Gamma_{Inactive}$ &&&&& 2B$_{1g}$ + B$_{2u}$ + E$_{2u}$ \\

$\Gamma_{Acoustical}$ &&&&& A$_{2u}$ + E$_{1u}$ \\

$\Gamma_{Raman}$ &&&&& 2E$_{2g}$ + A$_{1g}$ + E$_{1g}$ \\

$\Gamma_{Infrared}$ &&&&& A$_{2u}$ + E$_{1u}$ \\

\end{tabular}

%
%
%
%
%
%
%
%
%
%
%
%
%
%
%
%

\end{ruledtabular}
\end{table}


\begin{references}
\bibitem{2HNote}
Here, 2H is based on Ramsdell notation with 2 indicating the number
of S-W-S slabs within the unit cell and H representing the lattice
type (hexagonal).
\bibitem{optconst}
These optical constants include the real and imaginary parts of the
dielectric function, $\epsilon_1(\omega)$ and $\epsilon_2(\omega)$,
the energy loss function, $-$Im(1/$\epsilon$($\omega$)), and the
optical conductivity, $\sigma_1(\omega)$. In our analysis, the low
frequency data was extrapolated as a constant, and the high
frequency data was extrapolated with an exponent of 2.4.
\bibitem{conductivitynote}
In the optical conductivity, $\sigma_1(\omega)$, these features are
centered at 356.5 and 438.3 cm$^{-1}$ for the 2H compound and at
357.0 and 439.0 cm$^{-1}$ for the IF nanoparticles.
\bibitem{negativeL}
$\emph{e}_l^*$ cannot be obtained for the \emph{z}-polarized
interlayer A$_{2u}$ mode. The model for localized charge is
developed for studying only the intralayer E$_{1u}$ mode in the
layered 2H bulk material \cite{Burstein1970,Uchida1978}. Since the
Lorentz factor is negative for the A$_{2u}$ mode, no realistic
charge values could be determined.
\bibitem{jannote} Formally, the local site symmetry of a curved
sheet is C$_1$, and all A modes are active. But practically, the
nanoparticle curvature is small, so it is useful to consider the
additional spectral features as perturbations to the planar picture.
\bibitem{finestructure}
Within the clusters centered at $\sim$347, 379, and 426 cm$^{-1}$,
we observe vibrational fine structure at 338.7, 342.1, 345.0, 349.1,
355.09, 356.2; 362.6, 365.0, 368.1, 371.5, 376.0, 378.2, 381.9,
384.5, 388.0, 394.0, 396.8; and 420.4, 425.7, 430.1, 433.0
cm$^{-1}$, respectively.
\end{references}
\end{document}